\documentclass{article}

\usepackage[utf8]{inputenc}
\usepackage[margin=1in]{geometry}
\usepackage{hyperref}
\usepackage[T1]{fontenc}
\usepackage{titling}
\usepackage{pgfplots}
\usepackage{tikz}
\usepackage{float}
\usepackage{xspace}
\usepackage{etoolbox}
\usepackage{subcaption}
\usepackage{wrapfig}
\usepackage{authblk}
\usepackage{multirow}
\usepackage{array}
\usepackage{titlesec}
\usepackage[labelformat=simple]{subcaption}

\titlelabel{\thetitle.\quad}

  \usepackage[
    backend=biber,
    style=nature,
  ]{biblatex}
\hypersetup{
    colorlinks=true,
    linkcolor=black,
    filecolor=black,
     anchorcolor = black,
  citecolor = black,
    urlcolor=black,
}

\newcommand{\cut}[1]{}

\newcommand{\shortsection}[1]{\vspace{0.5em} \noindent{\bf #1}}
\newcolumntype{?}{!{\vrule width 1.5pt}}

\pgfplotsset{compat=1.17}

\title{VaultDB: A Real-World Pilot of Secure Multiparty Computation Within a Clinical Research Network}

\author[1]{\small Jennie Rogers}
\author[4]{\small Elizabeth Adetoro}
\author[1, 5]{\small Johes Bater}
\author[2]{\small Talia Canter}
\author[2]{\small Dong Fu}
\author[4]{\small Andrew Hamilton}
\author[4]{\small \\Amro Hassan}
\author[3]{\small  Ashley Martinez}
\author[4]{\small  Erick Michalski}
\author[2]{\small Vesna Mitrovic}
\author[4]{\small Fred Rachman}
\author[3]{\small Raj Shah}
\author[4]{\small \\Matt Sterling}
\author[3]{\small  Kyra VanDoren}
\author[2]{\small Theresa L. Walunas}
\author[1]{\small Xiao Wang}
\author[2]{\small Abel Kho}

\affil[1]{\footnotesize Northwestern University, McCormick School of Engineering}
\affil[2]{\footnotesize Northwestern University, Feinberg School of Medicine}
\affil[3]{\footnotesize Rush University}
\affil[4]{\footnotesize AllianceChicago}
\affil[5]{\footnotesize Duke University}

\date{\vspace{-1em}}

\begin{document}

\maketitle

\begin{abstract}


Electronic health records represent a rich and growing source of clinical data for research. Privacy, regulatory, and institutional concerns limit the speed and ease of sharing this data. VaultDB is a framework for securely computing SQL queries over private data from two or more sources. It evaluates queries using secure multiparty computation: cryptographic protocols that evaluate a function such that the only information revealed from running it is the query answer. We describe the development of a HIPAA-compliant version of VaultDB on the Chicago Area Patient Centered Outcomes Research Network (CAPriCORN).  This multi-institutional clinical research network spans the electronic health records of nearly 13M patients over hundreds of clinics and hospitals in the Chicago metropolitan area. Our results from deploying at three health systems within this network show its efficiency and scalability for distributed clinical research analyses without moving patient records from their site of origin.

\end{abstract}


\section{Introduction}
\label{sec:intro}

Data captured through routine clinical care in electronic health records (EHRs) are increasingly repurposed for clinical research with appropriate oversight. Moreover, pooling larger and larger patient cohorts across multiple institutions within data sharing networks enables researchers to conduct complex studies over larger and more diverse patient populations and creates improved opportunities to investigate rare disease. The ongoing COVID-19 pandemic has underscored speed and data sharing as necessities to generate timely insights for clinical and policy decisions. There are several challenges, however, to integrating data across multiple institutions for health research. In the US, the Health Insurance Portability and Accountability Act (HIPAA) of 1996~\cite{hipaa} defines how healthcare institutions manage, protect, and share patient data, as well as the penalties for data misuse. Moreover, the Health Information Technology for Econonomic and Clinical Heath (HITECH) Act of 2009~\cite{hitech} extended the entities subject to HIPAA requirements and introduced mechanisms to support more effective enforcement of the existing law.

As a result, investigators wishing to conduct research over EHRs from multiple healthcare institutions must establish data use agreements (DUAs) that outline how each institution, or data partner, maintains compliance with all applicable laws and regulations for their data. These agreements protect patient confidentiality.  They also specify the mechanisms for data sharing, the roles and responsibilities of all participants, and acceptable use of shared data. When a healthcare organization negotiates a DUA with a data network, it often takes months or more of committee and legal work. Hence, it may take many years for a network to build up a diverse cohort of patients for clinical research. Further complicating the situation, this research is subject to a witch’s brew of complex, partially overlapping regulations that vary by jurisdiction, divergent interpretations by stakeholders, and by the contents of the data itself. A second, related concern is building an ecosystem of support to encourage individuals and healthcare institutions to share their data for research without fear of reputation-damaging data breaches.

The Chicago Area Patient-Centered Outcomes Research Network (CAPriCORN) is clinical research network (CRN) made up of a coalition of healthcare and research institutions, patients, patient advocates, clinicians, community-based organizations (CBOs), and non-profit organizations committed to facilitating patient-centered clinical research through delivery of healthcare data. CAPriCORN’s mission is to develop, test, and implement policies and programs to improve health care quality, health outcomes, and health equity for the diverse population of the Chicago metropolitan region and beyond. Its members have a shared governance structure, a centralized IRB (Chicago Area IRB or CHAIRb) and engaged patients and community members as leaders within all aspects of network operations. Within the Chicago area, patients may seek care across multiple institutions.  Thus, CAPriCORN plays an essential role by linking these fragmented records together to form a more complete picture for research~\cite{kho2014capricorn, capricorn}.

Previously within CAPriCORN, we developed and demonstrated the use of privacy-preserving record linkage to address the challenge of deduplicating and aggregating patient data~\cite{kho2015design,mays2016evaluation,trick2021joining,walunas2017disease}. This method required use of an honest data broker or trusted third-party to act as an intermediary between the data partners and the data requester. In this role, the honest broker collects the relevant records from all sites, matches and joins records for the same patients, and returns these linked data to the investigator. CAPriCORN’s honest data broker is the Medical Research Analytics and Informatics Alliance (MRAIA)~\cite{mraia}. Because the honest broker handles data, even encrypted data, from multiple institutions its presence introduces some risks. If compromised, the honest broker leaks all data from all sites to an attacker. Even if the records remain encrypted, an attacker analyzing the network traffic between the honest broker and data partners might learn sensitive information such as the number of patients afflicted with a particular condition. Within CAPriCORN, participating institutions have signed DUAs that ensure strict data security and compliance processes, but this process has introduced significant delays that can hinder timely delivery of research data.

\subsection{Design and Deployment}
  \begin{wrapfigure}{r}{0.5\textwidth}
  \centering
  \includegraphics[width=0.5\textwidth]{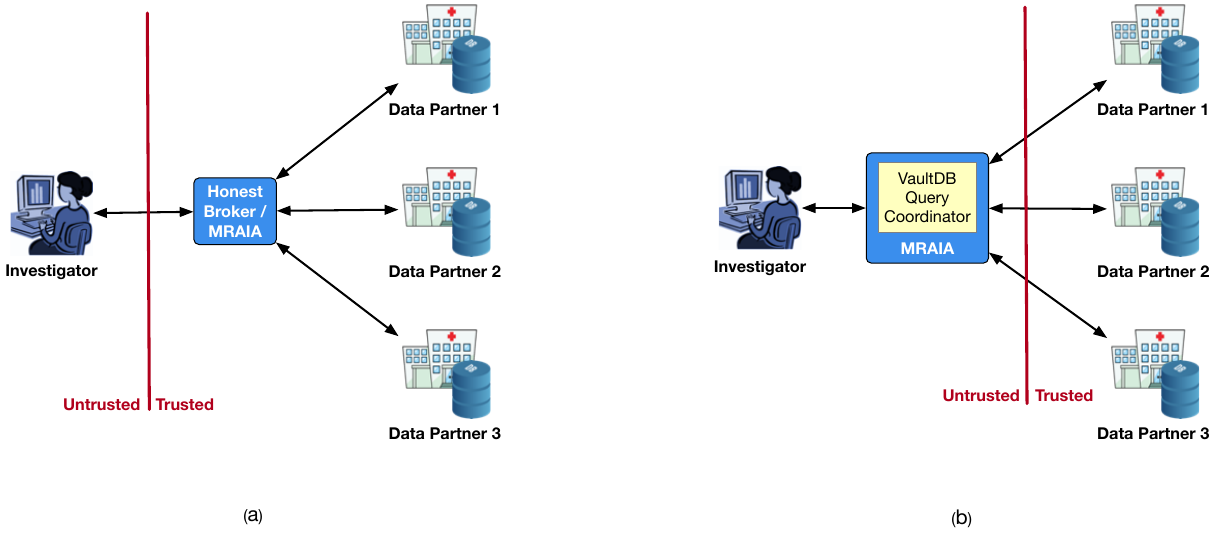}
  \caption{ Trust boundary for a CRN:  (a) with current best practices, and (b) in VaultDB}
    \label{fig:trust-boundary}
   \vspace{-1mm}
\end{wrapfigure}

Privacy-preserving analytics represent a significant advance for multi-institutional research, potentially obviating the need for an honest broker or DUAs. Figure~\ref{fig:trust-boundary} demonstrates how adding secure computation to the CRN reduces the risk associated with EHR sharing by making it so the data partners alone have access to their private records. In previous work, we verified the utility of SQL over  secure computation by running a representative workload of clinical research queries on de-identified EHR data in the HealthLNK Data Repository~\cite{Bater2017, galanter2013migration}. Since then, we have improved efficiency by integrating differential privacy~\cite{Bater2018}
and approximate query processing via input sampling~\cite{bater2020saqe} into this framework by selectively leaking information that is sufficiently fuzzy that it is provably impossible to result in data leakage. We now describe the design, security model, and real-world implementation and results of a secure computation across an active clinical research network with VaultDB, our prototype for computing CRN SQL queries securely.

VaultDB is our prototype for privacy-preserving analytics over a data federation such as a clinical research network (e.g., CAPriCORN). We call VaultDB a {\em private data federation} because, like a conventional data federation, it unifies multiple, independent databases for querying as if they were a single engine. To query the union of the datasets of all sites, an investigator, or client, submits SQL queries written against a common data model (CDM). All data providers support these shared table definitions, making the many databases appear as one to clients when they query the federation.   VaultDB integrates seamlessly with existing analytics frameworks because its interface mirrors that of a conventional data federation: it takes in a SQL statement from the analyst and returns a table of query results to them.

\begin{wrapfigure}{l}{0.5\textwidth}
  \begin{center}
  \vspace{-6mm}
    \includegraphics[width=0.48\textwidth]{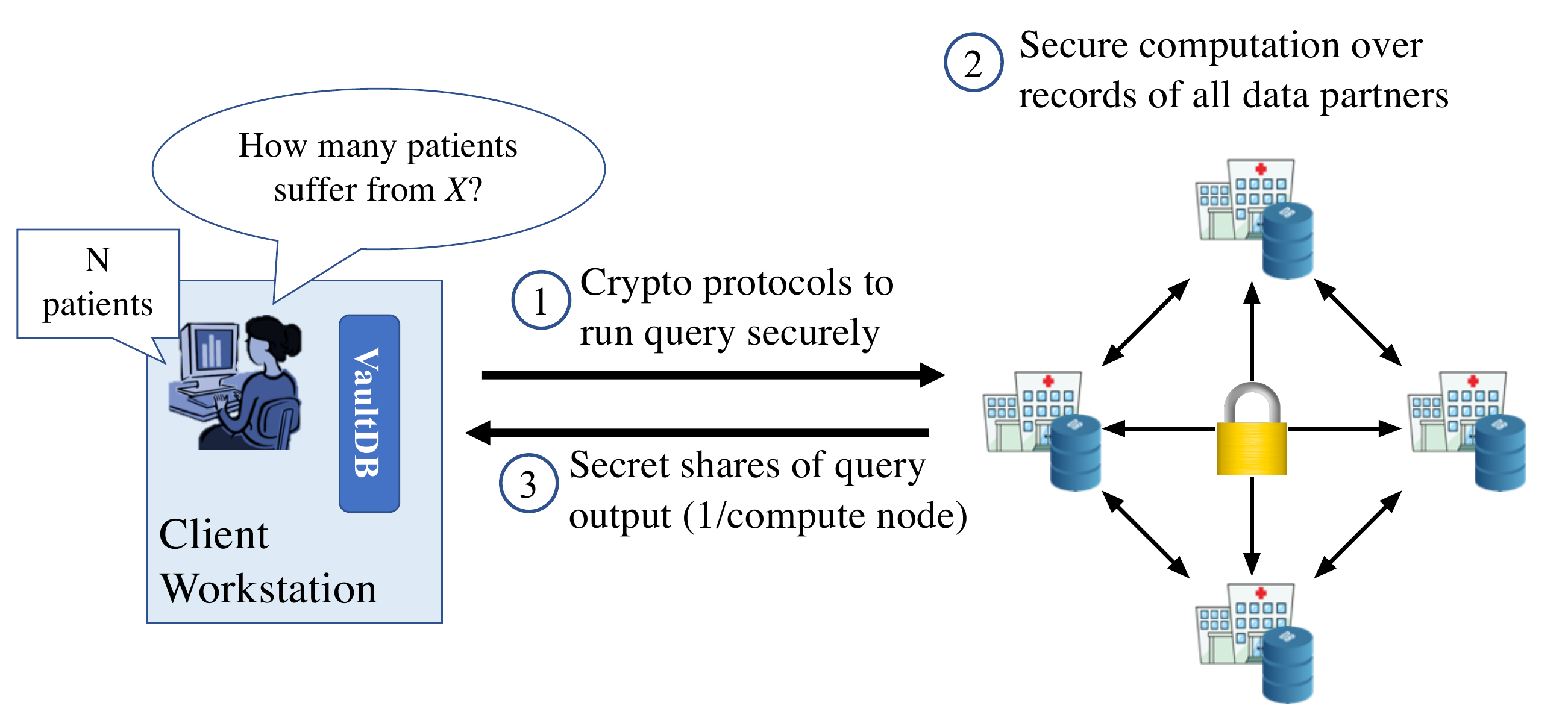}
  \end{center}
  \caption{VaultDB Workflow}
  \label{fig:workflow}
  \vspace{-2mm}
\end{wrapfigure}

Figure~\ref{fig:workflow} depicts VaultDB's query workflow.   Here, each healthcare site serves one of two roles. {\em Compute data partners} encrypt their data derived from EHRs for use in the study. In addition, they participate in the secure computation protocols VaultDB uses for query evaluation. Moreover, {\em data partners} encrypt their EHR data and upload them to computing nodes. When the analyst submits a query, $Q$, VaultDB translates $Q$ into cryptographic protocols that the data partners use to jointly compute it over their combined private inputs. With secure computation  the data partners  simulate the honest broker by passing messages to one another.  At the end of this secure protocol each compute data partner $i$ has $A_i$, their encrypted  share of the query answer.    Each computing data partner sends their $A_i$ to the analyst.   They assemble the shares, which may only be decrypted when taken together, and returns $A$ to the analyst.

 \shortsection{Secure Multi-Party Computation. } 
 Secure multi-party computation~\cite{4568207} is a subfield of cryptography that enables two or more parties to jointly compute a function over their private inputs without divulging them to others.  Here, the computing parties simulate a completely trustworthy, incorruptible third party by passing encrypted messages amongst themselves.  At the end of a query evaluation, the parties may combine their individual outputs to reveal the query answer.   Secure computation is quite expressive, it is Turing complete, therefore any function that is computable in conventional programs is also possible in this setting.

VaultDB’s cryptographic backend is powered by the Efficient Multi-Party (EMP) Toolkit~\cite{emp-toolkit}.  This open-source project was started in 2016 and is being actively maintained. The toolkit was developed to address two goals: 1) provide an easy way for researchers to build efficient cryptographic prototypes; 2) provide a user-friendly interface for non-cryptographic developers to write privacy-preserving applications.

\shortsection{Security Model and Deployment.}
For this study, we deploy EMP's two-party semi-honest protocols.   Hence, we trust the computing parties to faithfully execute the protocols as given with no collusion among them. Although each data partner agrees to not deviate from the protocol, they may attempt to deduce the private inputs of their peers by observing the secure computation and analyzing the secret shares. Secure multi-party computation provides cryptographically strong security guarantees for the data over which it computes.  This capability comes at a high performance cost since off-the-shelf algorithms usually have performance orders of magnitude slower than their security-free equivalents.  Section~\ref{sec:opt}  will describe a technique for reducing this overhead by decreasing the data over which we compute  securely.

VaultDB's secure query execution is {\em  oblivious}, or data-independent.  This means that the query's observable behavior does not change based on the contents of its inputs.  To make this possible, its execution times is proportional to the longest possible runtime since halting early may reveal information about the inputs that prompted this change in behavior.   This worst-case query evaluation prevents a curious computing node from deducing unauthorized information about its private input records.   Only the query answer $A$, and that which can be deduced from it, is revealed by participating in a private data federation query.    In existing systems, as shown in Figure~\ref{fig:trust-boundary}(a), 
 the honest broker  represents a single point of failure in our security model. In contrast, in VaultDB an adversary attacking an individual data partner cannot recover the query's inputs;  each one does not possess enough information to reconstruct them.  Moreover, since the query evaluation is oblivious an attacker gains nothing by intercepting network traffic between sites. The federation’s private inputs are not accessible outside their site of origin unless an attacker corrupts all computing parties.

To deploy VaultDB at healthcare institutions we needed to both uphold HIPAA requirements at each site and minimize this research prototype's privileges within these production environments. As such, we deployed the query processing engine as a Docker container on each site.  This setup was analogous to a virtual machine and made it possible to install the C++ program and its dependencies in a Linux environment without administrative privileges on mission-critical systems.   In addition, the containers had logging installed for audit trails to maintain compliance. These instances were deployed within the local network’s firewall in a ``demilitarized zone'' that limited their access to network resources such as file shares and database connectors. As a result of this design, VaultDB took all inputs as comma-separated value files rather than connecting to the local EHR datamart. To maintain compliance, all connections between data partners were encrypted using industry-standard encryption tools such as {\tt ssh}. Last, remote access to each site’s VaultDB instance required two-factor authentication.

\section{Methods}

We evaluated VaultDB with the following research goals:
\begin{itemize}
    \item {\bf Aim 1:} Replicate an existing CRN study protocol within VaultDB.
    \item {\bf Aim 2:} Extend this protocol to evaluate VaultDB’s ability to scale to larger inputs.
\end{itemize}

We started with an existing CAPriCORN project; the ElectroNic health Record-based dashboard to Improve Cardiovascular Health (ENRICH).  This study surveys patient data from healthcare organizations in the Chicago metropolitan area to identify demographic groups that may be suffering from untreated or under-treated hypertension. It was supported by the Illinois Department of Public Health (IDPH) through the Centers for Disease Control (CDC) and Prevention's DP18-1815 program, ``Improve the Health of Americans through Prevention of Diabetes, Heart Disease, and Stroke''~\cite{preventionDP1815}.  ENRICH was reviewed by CHAIRb (20190694124726-1) and received a non-human-subjects research determination.   We selected this study to evaluate VaultDB because it uses commonly available demographic and clinical data, spans multiple sites, and included large numbers of patient records. The original study analyzed patient data from the calendar year 2018, and our extended study covers 2018-2020, inclusive.

\begin{wraptable}{l}{7cm}
    \centering
       \vspace{-2mm}

    \begin{tabular}{|l|c|c|c|}\hline
         &  & Multi-Site  & \%  \\
        Site & Patients & Patients   & Overlap \\
        \hline
        AC & 31,165  & 3,140 & 10.1\%\\  
        NM  & 457,774 & 11,275 & 2.4\%\\
        RUMC  & 123,650 & 8,873 & 7.2\% \\\hline 
    \end{tabular}
    \caption{Study participants from each site.}
   \label{tbl:site_data}
   \vspace{-3mm}
\end{wraptable}

Three organizations participated in the VaultDB pilot: two large academic health systems, Northwestern Medicine (NM) and Rush University Medical Center (RUMC),  and a coalition of community health care organizations,  AllianceChicago (AC). This study was approved by CHAIRb (ID: 19110101). All data partners contributed EHRs to the pilot.  Table~\ref{tbl:site_data} shows the number of unique patients participating from each site. Multi-site patients contributed records at more than one site in at least one study year. This protocol spanned data from 600,000 unique patients over the three-year study period. A small fraction of the study population had fragmented care; they had records at multiple sites. Secure computation makes it possible for us to evaluate this protocol accurately without revealing the records of these multi-site participants to anyone.

   \begin{wrapfigure}{r}{0.5\textwidth}
  \begin{center}
  \vspace{-6mm}
    \includegraphics[width=0.48\textwidth]{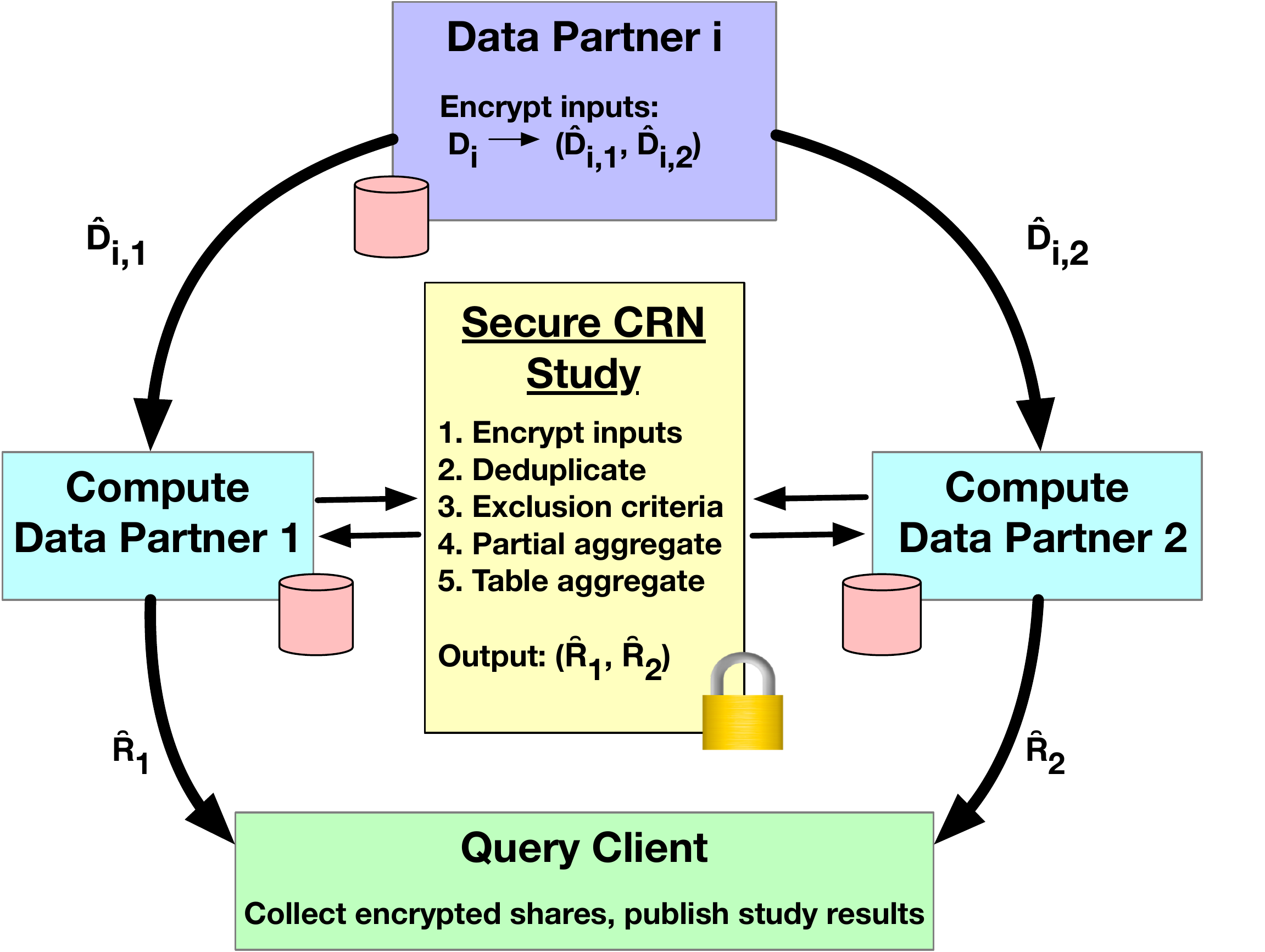}
  \end{center}
  \caption{VaultDB architecture.  All data partners contribute records to study.}
  \label{fig:arch}
  \vspace{-2mm}
\end{wrapfigure}

The ENRICH study accesses demographic data from the EHRs of two related cohorts: 1) individuals with a known diagnosis of hypertension (the denominator); and 2) patients with a diagnosis of hypertension and a blood pressure measurement greater than 140/90 during their most recent encounter at each healthcare site (the numerator). It stratifies study participants by age group, sex, race, and ethnicity and deduplicates patients with records at multiple sites. The study only included individuals seen in an ambulatory setting.  An individual is excluded from the study if they are deceased, recently pregnant, suffer from renal disease, had a recent kidney transplant, or had an inpatient encounter during the study period. If a patient matches the exclusion criteria at one study site, the protocol excludes all records of theirs, even if they are seen at multiple sites.

Figure~\ref{fig:arch} depicts VaultDB’s query evaluation workflow for ENRICH. Each site starts by regularizing its data, addressing any data quality issues, and reconciling it with the CRN’s CDM~\cite{ahmad2019challenges}. The data partners start with unique participant IDs for each patient in the study.  This is currently generated across CAPriCORN sites using Datavant’s tokenization software~\cite{datavant}. This step builds on our previous work on privacy-preserving record linkage~\cite{kho2015design}. In addition, each patient’s row is labeled as single-site or multi-site. All three sites encrypt their inputs with a secret sharing protocol. This protocol is equivalent to each holder of a secret share having an encrypted copy of the data but not the key with which to decrypt it.
 When data partner $i$ secret shares its data, $D_i$, it ``splits the secret'' and outputs $(\hat{D}_{i,1}, \hat{D}_{i,2})$. All sites complete this step.

Prior to evaluating a query, we assign two of the data partners as compute nodes that will perform the secure query evaluation.  We call them Alice and Bob for clarity.  Northwestern Medicine and Rush were the computing parties for this pilot. In practice, any combination of the data partners may take part in secure computation and, with additional cryptographic protocols, more than two may do so.  As a standard data partner, AllianceChicago prepares secret shares for their inputs and sends them to  Alice and Bob.   Alice  collects the ``1'' shares, $\hat{D}_{*,1}$, from all data partners and Bob receives all of the twos ($\hat{D}_{*,2}$).   This completes the setup process for the study.  Alice and Bob now pass encrypted messages amongst themselves to securely compute the queries in the ENRICH protocol.

The compute nodes first apply the study's exclusion criteria to winnow the data down to the records that qualify for the study.  It replaces the excluded records with dummies, or data placeholders that will not contribute to the query answer.  This ensures that the query’s evaluation remains oblivious. In the same pass over the data, the engine de-duplicates the patients, again replacing disqualified records with dummies. We compute this step as an aggregate. The oblivious aggregate in VaultDB first sorts the data by its group-by columns so that tuples that should be evaluated together are adjacent to one another and then does a linear scan over the sorted list to calculate the aggregates.

Next VaultDB computes a data cube~\cite{harinarayan1996implementing}, or multidimensional aggregate,  over all input rows. This jointly partitions the tuples over all four strata at once to compute fine-grained counts of the study variables. The output tuples are of the form  {\it (study\_year, age, sex, race,  ethnicity, numerator\_count, denominator\_count, numerator\_multisite\_count, denominator\_multisite\_count)}.  The output of this cube aggregation is revealed to no one in this study.  The query evaluator then repeatedly rolls up these partial counts to produce one table for each of the four demographics in the study.

\subsection{Optimizations}
\label{sec:opt} 
We introduced two optimizations in our query processing workflow to make VaultDB evaluate queries more efficiently and scale to larger datasets.  We describe each in turn below.

\shortsection{Multi-Site.}
In addition to the pipeline detailed above, we also evaluated VaultDB using the semi-join optimization introduced in our prior work~\cite{Bater2017}.  This is a generalization of a well-known relational database optimization.  Here, the data partners only compute securely over the tuples of patients that appear at multiple sites.   Each site also computes a  data cube over their single-site study participants' data and pad this result with dummy tuples to conceal how many  strata exist in this data.  They then secret share these aggregates using the protocol described earlier in this section.  After the compute nodes  evaluate the multi-site data cube, they securely add a single-site aggregate from each  data partner to it.  The rest of the protocol proceeds unmodified.

\shortsection{Batch.} In order to remain oblivious, these queries create very large intermediate result sizes.  To address this, we  evaluated  partitioning the data into smaller batches of tuples for processing rather than securely computing the protocol in one pass.  We divided up the tuples so that patients with fragmented care would be assigned to the same batch by hashing the patient IDs using a cryptographically secure hash function.  We then assign records to batches by taking the hash value modulus the batch count.   By processing the data incrementally, we scale to large data sets that otherwise would not fit into memory.  This makes the queries run more efficiently but requires additional secure computation to assemble the partial results from all batches.  This too generalizes on principles of distributed database systems.


\begin{table}[hbt!]
    \centering
 \resizebox{0.9\textwidth}{!}{

    \begin{tabular}{|l|cc|cc?cc|cc|}
    \hline
     & \multicolumn{4}{c?}{Aggregate Only}  &   \multicolumn{4}{c|}{Full Study Protocol} \\ 
    &   \multicolumn{2}{c|}{Patient Count}  & \multicolumn{2}{c?}{\% Fragmented}  & \multicolumn{2}{c|}{Patient Count}  & \multicolumn{2}{c|}{\% Fragmented} \\
     
   Stratification & Num. &Denom.  & Num. &  Denom. &     Num. &Denom.  & Num. &  Denom.\\
      \hline
Age (18-28) & 2,059 & 6,048 & 1.46\% & 1.80\% & 2,054 & 6,019 & 0.58\% & 0.45\%\\
Age (29-39) & 7,132 & 21,370 & 3.30\% & 3.08\% & 7,090 & 21,190 & 1.50\% & 0.83\%\\
Age (40-50) & 12,182 & 45,567 & 3.78\% & 3.35\% & 12,127 & 45,202 & 1.69\% & 0.80\%\\
Age (51-61) & 15,086 & 80,414 & 3.61\% & 3.17\% & 15,030 & 79,744 & 1.84\% & 0.83\%\\
Age (62-72) & 11,037 & 102,762 & 4.23\% & 2.66\% & 10,989 & 101,947 & 2.33\% & 0.80\%\\
Age (73-83) & 4,369 & 74,939 & 3.87\% & 2.37\% & 4,354 & 74,367 & 2.23\% & 0.77\%\\
Age (84-100) & 481 & 11,201 & 4.99\% & 2.37\% & 477 & 11,118 & 3.98\% & 0.75\%\\
\hline
Female & 24,143 & 173,922 & 3.85\% & 3.85\% & 24,026 & 172,513 & 2.01\% & 0.81\%\\
Male & 28,192 & 168,338 & 3.55\% & 3.55\% & 28,084 & 167,033 & 1.74\% & 0.78\%\\
\hline
Hispanic & 6,320 & 30,931 & 5.51\% & 5.51\% & 6,297 & 30,702 & 1.21\% & 0.74\%\\
Non-Hispanic & 43,116 & 255,070 & 3.43\% & 3.43\% & 42,915 & 252,611 & 2.08\% & 0.97\% \\
\hline

American Indian & 155 & 781 & * & 4.48\% & 155 & 779 & 0\% & *\\
Asian  & 1,652 & 9,946 & 2.66\% & 1.69\% & 1,649 & 9,904 & 1.33\% & 0.42\%\\
Black   & 10,466 & 48,221 & 5.24\% & 4.28\% & 10,383 & 47,510 & 2.96\% & 1.48\% \\
Native Hawaiian or  &  &  &  &   &  &  &  &   \\
 Pacific Islander  & 80 & 442 & * & 1.58\% & 80 & 442 & 0\% & 0\% \\
White  & 33235 & 208623 & 2.93\% & 2.36\% & 33,111 & 206,804 & 1.82\% & 0.88\% \\

\hline
   \end{tabular} }

    \caption{VaultDB 20202 ENRICH study results. Full study protocol has exclusion criteria and de-duplication.  }
    
    \label{tbl:clinical}
    \vspace{-2mm}
\end{table}

\section{Results}
\label{sec:results}

We evaluated VaultDB within  Docker containers deployed at NM and RUMC. The NM instance had 8 GB of RAM and the RUMC container had 11 GB of RAM.  Here, NM served as the generator of VaultDB's garbled circuits and RUMC evaluated them.  We used readily available hardware within the research infrastructure at each institution, without the need for specialized configurations. The two Chicago healthcare institutions are linked with a high-speed connection with bandwidth of  40 MB/sec.


Table~\ref{tbl:clinical} displays the clinical results from our study for the calendar year 2020.  Cell counts less than 11 were suppressed per the ENRICH study protocol to protect patient privacy. We evaluate the protocol in two ways.  The aggregate only experiment demonstrates the runtime of a streamlined version of the ENRICH protocol that only sums up the partial counts of patients where each site only inputs partial counts to secure computation. This represents a simplified data sharing protocol with no record linkage, another use case for CRNs. It takes in one data cube per site, each dummy-padded to the Cartesian product of the domains for the four study strata, adds them up, and generates the study outputs from this. Since this approach does not take into account the patients that are multi-site, aggregate only queries may report higher counts. Also, it lacks the fragmented care analysis of the original ENRICH protocol.  The full study protocol findings implement all five steps shown in Figure~\ref{fig:arch}.

Broadly speaking, both results follow similar distributions in the age and sex of their participants. Recall that study participants in the numerator had high blood pressure during their most recent measurement, while participants in the denominator had a known diagnosis of hypertension at the time of the study. The results imply that patients with poorly controlled blood pressure (numerator) are more likely to have fragmented care than patients with only a diagnosis of hypertension (denominator). Moreover, we see that older individuals in the numerator group are most likely to have fragmented care and that men and women have similar rates of care fragmentation in this study.  {\em VaultDB readily verified rates of uncontrolled hypertension in a large cohort of patients across three sites, stratified by important covariates of age, gender, race, and ethnicity.}

One unique aspect of the VaultDB approach is that it performs privacy-preserving suppression on results after running the study protocol over the union of the data of all sites.  In standard CRN protocols data partners are reluctant to contribute EHRs when their site does not have enough patients individually to produce cell counts that satisfy their privacy requirements because the honest broker would learn this private information. Since the EHRs are revealed to no one outside the site of origin, the private data federation is able to compute  aggregates where the data partners {\em jointly} satisfy the cell count requirements even when that is not possible for them individually.  This may enable research over traditionally under-served or rare disease sufferers without compromising healthcare best practices on privacy protection.

  \begin{figure}[hbt!]
\centering
\begin{subfigure}{0.47\textwidth}
  \centering
  \includegraphics[width=\textwidth]{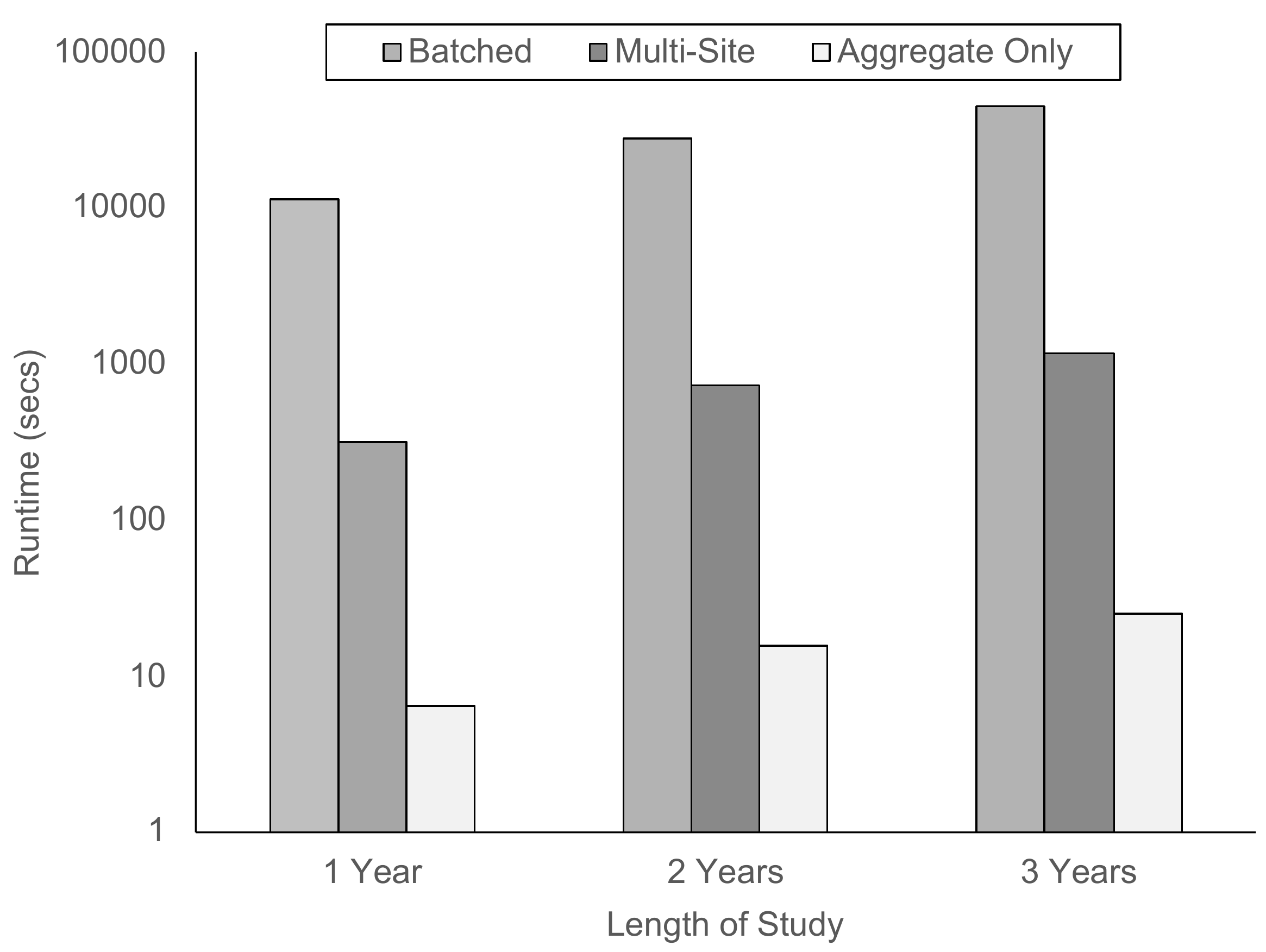}
  \caption{VaultDB query runtime with varied study length and use of secure computation.}
  \label{fig:vdb-runtime}
\end{subfigure}%
\hfill 
\begin{subfigure}{0.47\textwidth}
  \centering
  \includegraphics[width=\textwidth]{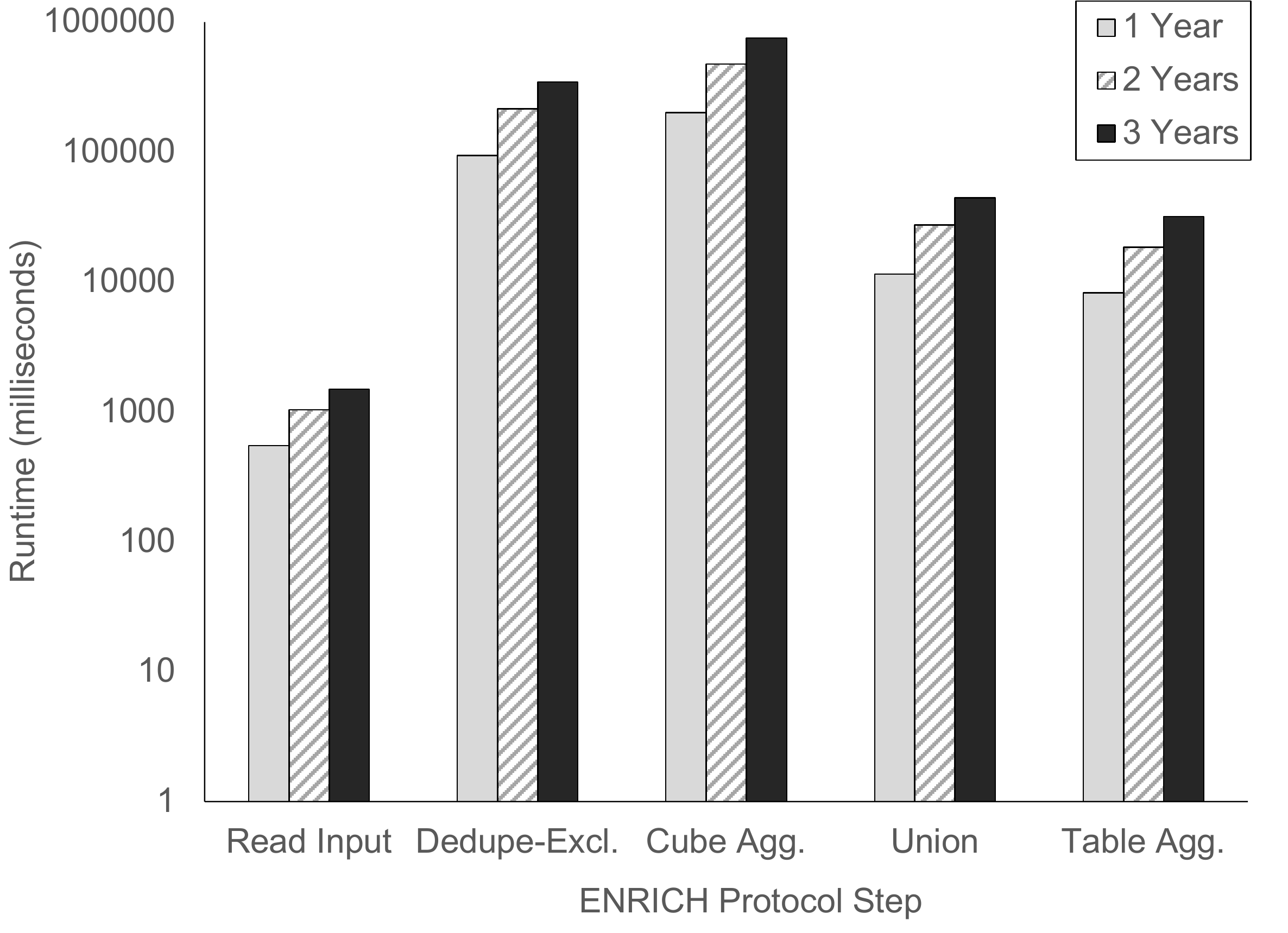}
  \caption{Runtime of each step in ENRICH  study protocol scaling from one to three years of data.}
  \label{fig:vdb-steps}
\end{subfigure}
\caption{VaultDB performance on ENRICH study protocol.}
\label{fig:vdb-results}
\vspace{-4mm}
\end{figure}

\subsection{VaultDB Performance}



Figure~\ref{fig:vdb-results} depicts our evaluation of VaultDB's performance for the ENRICH study.   We evaluate VaultDB's secure query evaluation in three ways.  First, the  batched evaluation securely computes the ENRICH workflow--implementing all five steps in Figure~\ref{fig:arch}.  It produced the results in Table~\ref{tbl:clinical}.   The number of rows is higher than our patient counts because if a patient participates in the study over multiple years they contribute one row per year.   This system computes on up to 1M input tuples.   Our optimizations are described in Section~\ref{sec:opt}. To scale to these larger datasets, we partitioned the data into  25 batches in our experiments.  We also evaluated increasing or decreasing the batch size, and these results are omitted due to space constraints.  Our choice of $n=25$ for batching resulted in a profitable trade-off of minimizing the cost of assembling the output of each batch processing while providing efficient performance by enabling the secret shares of each batch to fit into memory.    The second query evaluation,  multi-site, implements the same protocol but only computes securely on the EHRs of patients with fractured care.  The last one, aggregate-only, is described above.

In Figure ~\ref{fig:vdb-runtime} we evaluated incrementally scaling up this workflow one year at a time.  Note the logarithmic y-axes in both figures.  The first series covers the first study year, the second the first two years, and finally the last covers all years.  Table~\ref{tbl:input_sizes} shows the study's input size for each year.

The multi-site experiments demonstrate the performance of VaultDB when it performs deduplication and applies the exclusion criteria in secure computation only for the patients with fragmented care.  We compute these steps on about 3\% of the records in our experiments.  Since all other counts are computed at the data partners locally, we add a step between the data cube aggregate and the output aggregates to incorporate these single-site partial counts into the data cube.  

  \begin{wraptable}{r}{6cm}
    \centering
    \vspace{-4mm}
    \begin{tabular}{|l|c|c|}\hline
        Study  & Input  & Multi-Site    \\ 
          Length &  Rows &  Rows  \\
        
        \hline
        1 Year &  317,323  & 9,659  \\  
        2 Years  & 666,402 &  18,596 \\
        3  Years & 1,020,470  &  28,043 \\
        \hline
    \end{tabular}
    \caption{Input size vs  years of study.}
    \vspace{-1.5mm}
   \label{tbl:input_sizes} 
\end{wraptable}

 We see the aggregate-only analysis gradually increases in runtime (from about 5 seconds to 15 seconds) as the domain of each site’s inputs increases. Here, the number of valid years goes up as we include more study years. The fast runtimes of aggregate-only analysis make it imminently practical for large datasets. This result demonstrates a real-world practice (simple aggregation of study variables) in which we may trade accuracy for efficiency in private data federation queries. The results demonstrate that {\em all three query evaluation strategies scale efficiently} as we add more data to the processing pipeline.

We examine this increase in runtime as the data grows in Figure~\ref{fig:vdb-steps}. It shows the results of the multi-site optimized experiments for each step in  ENRICH. We see that the first ones, data ingest through data cube, grow in proportion to the input size. Secret sharing the input records requires network transmission, therefore, this cost naturally grows as we input more data. The data preparation steps (deduplication, applying the exclusion criteria, and computing the data cube) have two sorts. Each oblivious sort has a cost of $O(n \log^2 n)$, where $n$ is the count of input tuples.  Hence, their cost grows more quickly than steps that are  linear time such as adding up the single-site partial counts when unioning for the semi-join optimization. The latter’s cost is proportional to the maximum size of the data cube, a few hundred rows per study year.




\section{Discussion}
\label{sec:conclusions}

In this work, we describe the design, practical real-world health application, and performance of VaultDB, a private data federation.  The goal of this research is to facilitate clinical research over diverse patient populations to enable faster,  accurate, and cryptographically secure, health research and insights for public policy decisions. We aim to do this in settings where conventional data sharing agreements are either  impossible or too onerous to produce actionable insights in a timely fashion. Secure querying of EHRs will enable researchers and clinicians to efficiently conduct health research or public health surveillance even if the relevant patient records are fragmented over multiple sites. Understanding  fragmented care is especially important to serving more  vulnerable community members such as individuals experiencing homelessness, mental health challenges, or socio-economic issues.  	Our approach builds on our prior work in privacy-preserving linkage and analyses.  Responding to real-world needs, VaultDB makes substantial, potentially game-changing improvements to the CRN security model.  It deploys state-of-the-art cryptographic protocols to reducing the need for an honest broker thereby greatly reducing the network's exposure to  potential threats.  Moreover, it does so for SQL queries making this technology accessible to a broad audience of analysts without formal training in cryptography.

In this real-world demonstration, we used containerized software installed on readily available hardware inside three institutions and our system performed efficiently and scaled to moderate-sized workloads.  Our results demonstrate efficient evaluation of a complex clinical research study protocol without revealing the individual health records to anyone outside the site of origin.  This may transform how we engage in data sharing in clinical research in the future.  
Our ongoing work is focused on further integration of VaultDB with researcher and public health workflows, improving the ease for translating common research queries into secure multiparty compute ones, and further enhancing query efficiency.  


\section{Acknowledgments}
This work was supported in part by NSF Award \#1846447 and \#2016240, and CDC Grant 13286008I\/1NU58DP006511-03-00.  We thank the teams at MRAIA, CAPriCORN, and CHAIRb for their support during this study.


\section{Conflicts Statement}

The authors have no conflicts of interest to report regarding their work on this study.

\newpage

 \printbibliography

@INPROCEEDINGS{4568207,
  author={Yao, Andrew Chi-Chih},
  booktitle={27th Annual Symposium on Foundations of Computer Science (FOCS 1986)}, 
  title={How to generate and exchange secrets}, 
  year={1986},
  volume={},
  number={},
  pages={162-167},
  doi={10.1109/SFCS.1986.25}}

@article{Bater2017,
author = {Bater, Johes and Elliott, Gregory and Eggen, Craig and Goel, Satyender and Kho, Abel and Rogers, Jennie},
doi = {10.14778/3055330.3055334},
issn = {21508097},
journal = {PVLDB},
number = {6},
pages = {673--684},
title = {{SMCQL: secure querying for federated databases}},
volume = {10},
year = {2017}
}

@article{Bater2018,
  title={Shrinkwrap: efficient SQL Query Processing in Differentially Private Data Federations},
  author={Bater, Johes and He, Xi and Ehrich, William and Machanavajjhala, Ashwin and Rogers, Jennie},
  journal={PVLDB},
  volume={12},
  number={3},
  pages={307--320},
  year={2018},
  publisher={VLDB Endowment}
}

@article{bater2020saqe,
  title={{SAQE}: practical privacy-preserving approximate query processing for data federations},
  author={Bater, Johes and Park, Yongjoo and He, Xi and Wang, Xiao and Rogers, Jennie},
  journal={Proceedings of the VLDB Endowment},
  volume={13},
  number={12},
  pages={2691--2705},
  year={2020},
  publisher={VLDB Endowment}
}

@misc{emp-toolkit,
      author = {Xiao Wang and Alex J. Malozemoff and Jonathan Katz},
      title = {{EMP-toolkit: Efficient Multiparty computation toolkit}},
      howpublished = {\url{https://github.com/emp-toolkit}},
      year={2016}
    }

@article{kho2015design,
  title={Design and implementation of a privacy preserving electronic health record linkage tool in Chicago},
  author={Kho, Abel N and Cashy, John P and Jackson, Kathryn L and Pah, Adam R and Goel, Satyender and Boehnke, J{\"o}rn and Humphries, John Eric and Kominers, Scott Duke and Hota, Bala N and Sims, Shannon A and others},
  journal={Journal of the American Medical Informatics Association},
  volume={22},
  number={5},
  pages={1072--1080},
  year={2015},
  publisher={Oxford University Press}
}

@article{kho2014capricorn,
  title={CAPriCORN: Chicago area patient-centered outcomes research network},
  author={Kho, Abel N and Hynes, Denise M and Goel, Satyender and Solomonides, Anthony E and Price, Ron and Hota, Bala and Sims, Shannon A and Bahroos, Neil and Angulo, Francisco and Trick, William E and others},
  journal={Journal of the American Medical Informatics Association},
  volume={21},
  number={4},
  pages={607--611},
  year={2014},
  publisher={BMJ Publishing Group}
}

@misc{trick2021joining,
  title={Joining Health Care and Homeless Data Systems Using Privacy-Preserving Record-Linkage Software},
  author={Trick, William E and Hill, Jennifer C and Toepfer, Peter and Rachman, Fred and Horwitz, Beth and Kho, Abel},
  journal={American journal of public health},
  volume={111},
  number={8},
  pages={1400--1403},
  year={2021},
  publisher={American Public Health Association}
}

@article{mays2016evaluation,
  title={An evaluation of recurrent diabetic ketoacidosis, fragmentation of care, and mortality across Chicago, Illinois},
  author={Mays, James A and Jackson, Kathryn L and Derby, Teresa A and Behrens, Jess J and Goel, Satyender and Molitch, Mark E and Kho, Abel N and Wallia, Amisha},
  journal={Diabetes Care},
  volume={39},
  number={10},
  pages={1671--1676},
  year={2016},
  publisher={American Diabetes Association}
}

@article{walunas2017disease,
  title={Disease outcomes and care fragmentation among patients with systemic lupus erythematosus},
  author={Walunas, Theresa L and Jackson, Kathryn L and Chung, Anh H and Mancera-Cuevas, Karen A and Erickson, Daniel L and Ramsey-Goldman, Rosalind and Kho, Abel},
  journal={Arthritis care \& research},
  volume={69},
  number={9},
  pages={1369--1376},
  year={2017},
  publisher={Wiley Online Library}
}

@article{ahmad2019challenges,
  title={Challenges to electronic clinical quality measurement using third-party platforms in primary care practices: the healthy hearts in the heartland experience},
  author={Ahmad, Faraz S and Rasmussen, Luke V and Persell, Stephen D and Richardson, Joshua E and Liss, David T and Kenly, Pauline and Chung, Isabel and French, Dustin D and Walunas, Theresa L and Schriever, Andy and others},
  journal={JAMIA open},
  volume={2},
  number={4},
  pages={423--428},
  year={2019},
  publisher={Oxford University Press}
}

@article{harinarayan1996implementing,
  title={Implementing data cubes efficiently},
  author={Harinarayan, Venky and Rajaraman, Anand and Ullman, Jeffrey D},
  journal={Acm Sigmod Record},
  volume={25},
  number={2},
  pages={205--216},
  year={1996},
  publisher={ACM New York, NY, USA}
}

@article{galanter2013migration,
  title={Migration of patients between five urban teaching hospitals in Chicago},
  author={Galanter, William L and Applebaum, Andrew and Boddipalli, Viveka and Kho, Abel and Lin, Michael and Meltzer, David and Roberts, Anna and Trick, Bill and Walton, Surrey M and Lambert, Bruce L},
  journal={Journal of medical systems},
  volume={37},
  number={2},
  pages={1--8},
  year={2013},
  publisher={Springer}
}

@misc{datavant,
  title = {{Datavant: Connecting Health Data to Improve Patient Outcomes}},
  howpublished = {\url{https://datavant.com}},
  note = {Accessed: 2022-02-25}
}

@misc{capricorn,
author = {{Chicago Area Patient-Centered Outcomes Research Network}},
howpublished = {\url{https://www.capricorncdrn.org/about/}},
  note = {Accessed: 2022-02-25}
}

@misc{mraia,
author={{Medical Research Analytics and Informatics Alliance}},
title = {Public health and research informatics},
howpublished = {\url{https://www.mraia.org}},
  note = {Accessed: 2022-02-25}
}

@misc{hipaa,
  author = {{Centers for Medicare \& Medicaid Services}},
  title = {{The Health Insurance Portability and Accountability Act of 1996 (HIPAA)}},
  year = 1996
}

@article{hitech,
  title={Launching {HITECH}},
  author={Blumenthal, David},
  journal={New England Journal of Medicine},
  volume={362},
  number={5},
  pages={382--385},
  year={2010}
}

@misc{preventionDP1815,
author={{Centers for Disease Control and Prevention}},
title={{DP18-1815}: Improving the Health of Americans Through Prevention and Management of Diabetes, Heart Disease, and Stroke},
 howpublished = {\url{https://www.cdc.gov/diabetes/programs/stateandlocal/funded-programs/dp18-1815.html}}
 }

\end{document}